\documentclass[preprint,showpacs,aps,prb]{revtex4}
\usepackage{graphicx}

\begin{document}
\bibliographystyle{prsty}

\title{Transport of magnetic vortices by surface acoustic waves}
\author{Fabian Jachmann and Carsten Hucho\footnote{hucho@pdi-berlin.de}}

\affiliation{Paul-Drude-Institut f\"ur Festk\"orperelektronik, Hausvogteiplatz 5-7, 10117 Berlin, Germany}

\begin{abstract}
In a thin film of superconducting YBa$_2$Cu$_3$O$_7$ the impact of surface acoustic waves (SAWs) traveling on the piezoelectric substrate is investigated. A pronounced interaction between the ultrasonic waves and the vortex system in the type II superconductor is observed. The occurrence of a SAW-induced dc voltage perpendicular to the sound path is interpreted as {\em dynamic pinning} of vortices by the piezoacoustic SAW, which acts as a conveyor for the fluxquanta. Its antisymmetry with respect to the magnetic field directly evidences the induced, directed flux motion.This dynamic manipulation of vortices can be seen as an important step towards flux-based electronic devices.

\end{abstract}

\maketitle
\section{introduction} The pinning strength of magnetic vortices is crucially determining the critical current density, $j_{c}$, in the superconducting state of type-II superconductors.
While $j_{c}$ is low as long as the magnetic vortices
are able to move in the presence of a driving force, $j_{c}$ increases with
increasing pinning of the vortices \cite{Blatter94}. In order
to enhance the critical current density for technical applications, pinning is usually introduced
by fabricating static non-superconducting structures in the material. 
In this context the effect of matching between the regular vortice-lattice and
static artificial pinsite-lattice structures is widely investigated \cite{Daldini74,Civale91}. A
pronounced increase of $j_c$ occurs at the fields where vortex-structure and pinsite-structure coincide. The ability to modify the pinning structure dynamically by simply changing
external parameters in real time would clearly lead to exciting consequences: from adaptation
of the pinsite-density to the applied magnetic field, thereby tuning of the 
critical current density dynamically (applicable in tunable fuses and switches) to the 
targeted manipulation of the local magnetic flux density and transportation of 
individual vortices.\\
Since the influence of strain on the superconducting transition temperature is well
known \cite{Pickett97} and since the electric field effect on T$_c$ \cite{Mannhart91,Frey95} and 
associated significant changes in the pinning \cite{Mannhart91a} in high-T$_c$ superconductors are well 
established, a surface acoustic wave (SAW) traveling on a piezoelectric substrate 
locally modifies the superconducting properties of the high-T$_c$ superconducting film
by the combined strain- and electric-field. This will result in a time-dependent, moving
'ripple'-structure of spatially alternating weak and strong pinning-regimes, with the wavelength 
of the SAW, which exerts a drag on the magnetic vortex system. The interaction between the
SAW and the vortex ensemble is believed to be due to the modulation of the superconducting
properties of the film and an interaction of the piezoelectric field with the charge carriers
and is therefore different from the interaction of SAW and vortices
via the so-called acoustoelectric effect \cite{Gutliansky94, Sonin96}.
The acoustoelectric effect as calculated for an YBCO film by Gutlianskij considers only {\em sound induced} currents
as driving force for the vortices. The electrons, however, are essentially clamped to the ions at low frequencies
and low magnetic fields, leading to a vanishing net ac-electric current. 
Here we show direct evidence for the dragging of vortices in a well defined direction by a
SAW on a piezoelectric substrate.\\
This mechanism to move vortices in a well defined manner by an easily accessible external parameter
can in principle be used to manipulate the magnetic flux-density locally
(by using focusing acoustic beams) and to manipulate individual vortices and
vortex-batches (with the possibility to 'handle' vortices and antivortices in 
close proximity). While these processes constitute interesting physical phenomena
on their own, a number of technological consequences can
be thought of, such as possible new ways for data representation and
possible drives for magnetic microparticles.

\section{samples and techniques}
\begin{figure}
\centerline{\includegraphics*[width=8cm]{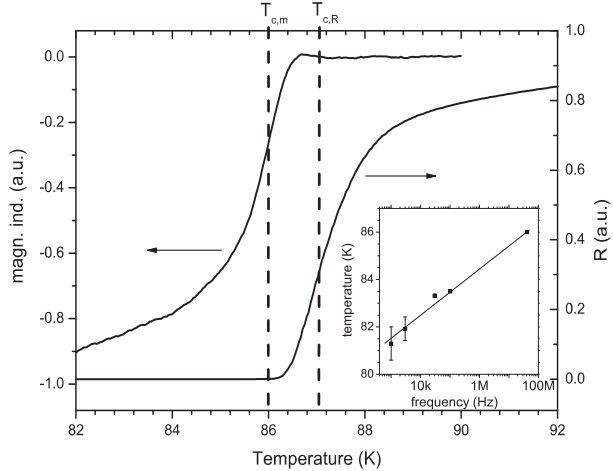}}
\caption{Temperature dependent dc resistivity at zero applied field and real part of the ac induction
at 42 MHz (smoothed). The magnetic (T$_{c,m}$) and resistive (T$_{c,\rm R}$) transition temperatures were defined as the inflection
points of the temperature dependent curves and are marked by vertical dashed lines. The inset shows the
frequency-dependence of T$_{c,m}$.} \label{superconductivity}
\end{figure}

100nm thick c-axis oriented YBCO films where grown on polished y-cut LiNbO$_3$ wafers by
laser-ablation (THEVA). Thin strips (0.5 $\times$ 4 mm$^2$) were masked and etched for 4-probe
resistivity experiments and gold contacts were manufactured in the
process. SAW interdigital transducers with a base
frequency of 42MHz were photolithographically produced perpendicular 
to the current path. The superconducting properties of the films were checked with four-probe 
resistivity and with ac-susceptibility measurements. The magnetic (T$_{c,m}$) and resistive 
(T$_{c,\rm R}$) transition temperatures were defined as the inflection points of the temperature 
dependent curves. The sample shows a clear, sharp transition into the
superconducting state (see figure \ref{superconductivity}). 
As usual, the resistively determined superconducting transition appears at significantly
higher temperatures than the transition as determined from the peak in the imaginary
part of the ac-susceptibility.
This reflects the different intrinsic properties probed: the resistive transition
marks the percolation temperature for a path of vanishing resistivity, while
ac-susceptibility experiments measure volume effects of screening currents and the
peak position is seen as the transition temperature into the glass phase.
$T_{c,m}$ shows an exponential frequency dependence over a range of 4 orders of magnitude, as expected 
\cite{geshkenbein91}. As argued by Geshkenbein et al. \cite{geshkenbein91} the peak
in the imaginary part of the ac-susceptibility approaches the true glass transition only
at the (theoretical) limit of zero frequency but shifts to higher temperature with
increasing frequency.
In order to be able to compare the SAW related results with the magnetic transition, the
magnetic-induction-experiment presented in figure \ref{superconductivity} was performed at
the same frequency as the acoustic experiments discussed here, namely 42~MHz.
\\
\section{SAW induced dc voltage}
\begin{figure}
\centerline{\includegraphics*[width=9cm]{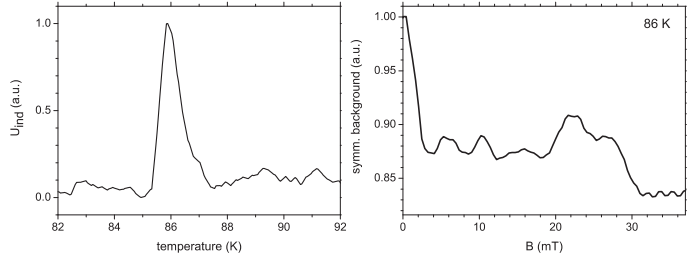}}
\caption{SAW-induced dc voltage (background peak) at zero applied magnetic field (left). Field dependence of the symmetric background-peak 
amplitude (right)}\label{dc_background}
\end{figure}

The aim of the work presented here is to demonstrate the interaction of a traveling
SAW with the magnetic vortex ensemble. The experiment is designed
to detect the interaction of the vortices with the SAW-related traveling 'pinning-grid' by measuring a 
quantity that can be clearly attributed to a resulting vortex-drag. This is in contrast
to the work of Pankert et al., who studied the pinning-depinning transition of vortices to {\em static} defect-sites
{\em in the crystal-lattice} \cite{Pankert90, Pankert90a}.\\
A SAW on the piezoelectric LiNbO$_3$ substrate is comprised of
a traveling strain wave, accompanied by an alternating electric field (with the 
same wavelength as the soundwave) and the remnants of the rf-electromagnetic 
background radiating from the transducer-structure and the leads. This combined signal
interacts with the vortex ensemble via a number of mechanisms, most of which are isotropic. 
The hallmark of components related to the traveling soundwave is its directionality: 
Only effects related to the directed motion of the vortex-ensemble induced by the SAW should 
consequently show a sign-reversal by inverting the static applied magnetic field.\\
\begin{figure}
\centerline{\includegraphics*[width=8cm]{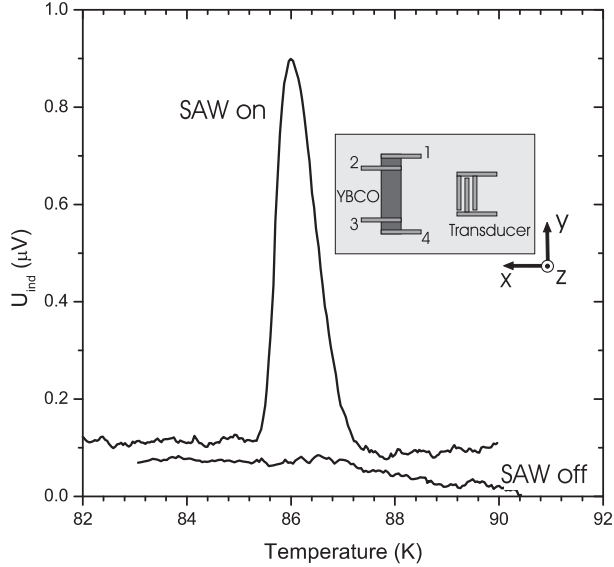}}
\caption{Induced dc voltage between contacts 2 and 3 with and without surface acoustic wave at an applied field
of -6 mT. The SAW propagates along x, the magnetic field is applied along z.}\label{ac_induced}
\end{figure}
Figure \ref{ac_induced} shows the temperature dependence of the dc-voltage drop, U$_{ind}$ between contacts 2 and 3 
of the superconducting film at zero applied current for an applied magnetic field of -6 mT (along z).
As expected for a superconductor in the current free configuration, there are no significant features 
in the signal when sweeping the temperature through the superconducting transition. 
However, a pronounced dc-voltage peak appears in the vicinity of T$_{c,m}$ when the superconducting film 
is simultaneously stimulated by a surface acoustic wave (42MHz) traveling along x, i.e. perpendicular to the electrical path.
This effect of a SAW on the electric behavior of the superconducting film is
very pronounced and it is intriguing to attribute the induced dc voltage to a SAW-driven
directed motion of the vortex ensemble in the background of superconducting charge carriers.\\
A similar peak, however, is observed also in zero applied magnetic field (see figure \ref{dc_background}). While this peak dominates
the dc-signal, it has clearly nothing to do with SAW-vortex interaction (it will be attributed to an ac-dc 
conversion-effect described below). This background dc-peak persists as a symmetric (with respect to the magnetic field)
contribution to the SAW-induced dc signal at nonzero fields. The normalized amplitude of this background peak is plotted vs. B in figure \ref{dc_background}, right panel. The field dependence is clearly non-monotonous. At zero field the peak-height is taken 
directly from the zerofield experimental data. At higher magnetic field strength the temperature-dependent dc-voltage measurements 
for positive and negative applied magnetic field were averaged.\\

\begin{figure}
\centerline{\includegraphics*[width=9.8cm]{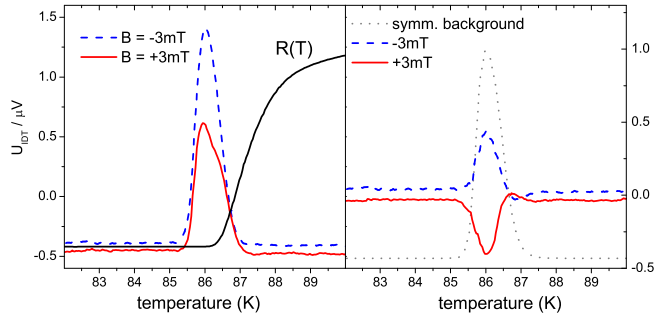}}
\caption{SAW-induced dc voltage in the presence of a magnetic field of +3mT and -3mT, respectively. The left
panel shows the raw-data and the resistive transition. In the right panel the background (which is symmetric
with respect to magnetic field) is displayed as well as the separated antisymmetric contributions
to the dc-voltage.}\label{plus_minus}
\end{figure}

\begin{figure}
\centerline{\includegraphics*[width=8.9cm]{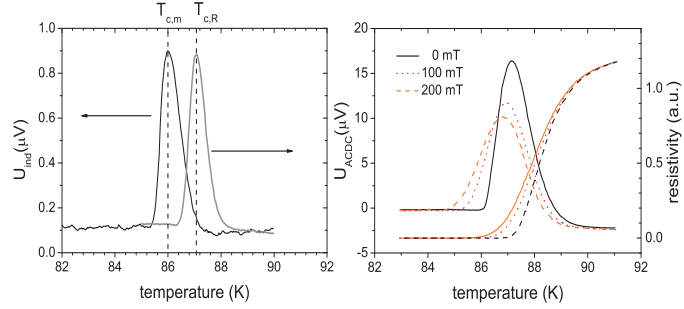}}
\caption{Temperature dependent ac-induced dc voltage in the presence of a magnetic field of -6mT, 
for SAW- (low-T peak, U$_{ind}$) and direct-ac-current excitation (high-T peak, U$_{ACDC}$) (left). Temperature dependent direct ac-induced dc voltage and resistivity at different applied magnetic field strengths in a similar sample (right).}\label{magnetic_electric_A}
\end{figure}

The quantity related to directed vortex-motion, U$_\Phi$, should be antisymmetric with respect to the applied magnetic field. 
We therefore separate the symmetric and antisymmetric contributions by repeating each temperature dependent
measurement for positive and negative applied field and subtracting the average of both runs from the data.
A typical result is shown for $\pm$ 3 mT in figure \ref{plus_minus}.
While the left panel shows the raw-data, the separated signals are shown in the right panel. These
data were obtained by subtracting the symmetric background, which is included as dotted line. 
The antisymmetric dc voltage, $V_\Phi$,
is ascribed to the SAW-induced directed motion of vortices. With increasing magnetic field
strength the feature shifts slightly to lower temperatures and becomes hardly detectable
for fields above 30 mT. Since the SAW-related pinning-grid moves with the velocity
of sound ($\sim 3500m/s$), the coupling of vortices to this dynamic pinning structure
depends strongly on the pronounced temperature- and field-dependence of the vortex-mobility (see, e.g., \cite{Doettinger94}).
The effect of vortex-drag is, therefore, expected to be most prominent in a very narrow temperature and
field range at the boundary between the vortex-glass phase and the thermally activated flux flow regime. 
Furthermore, the presence of a high density of static, isotropic pinsites in the YBCO film may actually 
enhance the observed ac-dc conversion effect. By impressing a directional bias on the otherwise isotropic 
pinning potential of the intrinsic pinning sites, switching on the SAW switches on the so-called ratchet-effect, 
which was observed in high T$_c$ superconducting films with artificially fabricated static, asymmetric pinning 
centers \cite{Villegas03, Woerdenweber04, Vondel05} - thereby leading to the rectification of vortex motion.
\\
In the following the contribution of the background to the observed dc-voltage will be discussed in
more detail. This feature is per definitionem independent of the sign of the applied magnetic field. It can
be the result of a number of interactions, first, the rf-electromagnetic background, second, modulation
of vortices pinned to the crystal lattice, or third, local heating.\\
First, due to its huge wavelength as compared to sample size, the rf-electromagnetic 
background can be seen as an isotropic, time-dependent modulation of the {\em applied} flux density, \.{B},
which results in induced ac-electric screening currents. At the resistive transition, T$_{c,R}$,
the current ($I$) voltage ($U$) characteristic is highly nonlinear and $U\propto \exp(-A/I^a)$ in the 
vortex-glass regime\cite{Blatter94}. Inserting an rf-current (I$_\sim$) with frequency $\omega$
\begin{equation}\label{harmonicCurrent}
    I_\sim=I_0 sin(\omega t)
\end{equation}
into a series expansion of this nonlinear current-voltage characteristic 
\begin{equation}\label{series_expansion}
    U(I)=U(0)+\frac{dU}{dI}\mid _0 I+
    \frac{1}{2!}\frac{d^2U}{dI^2}\mid _0 I^2+\frac{1}{3!}\frac{d^3U}{dI^3}\mid _0
    I^3,
\end{equation}
clearly results in a dc-component of the voltage
\begin{equation}\label{VDC}
    U_{DC}=\frac{dU^2}{dI^2}\mid _0 \left(\frac{I_\sim^2}{4}\right).
\end{equation}
\begin{sloppypar}
Ikegawa et al. \cite{Ikegawa88} analyzed the induced dc-voltage by applying an ac current
to a BaPb$_{1-x}$Bi$_x$O$_3$ sample. They clearly observed the ac-induced dc voltage
and showed that, interestingly, also higher order terms contribute to the induced dc voltage. 
A detailed analysis of the background ac-dc-conversion is not attempted here.\\
In order to support the role of the ac-dc-conversion-effect for the background dc signal, 
an ac-voltage of 42~MHz was applied directly to the superconducting film (contacts 1 and 4, see figure \ref{ac_induced}), 
while the temperature dependent dc-voltage (contacts 2 and 3) was measured. Indeed, a pronounced dc-voltage 
peak is clearly observed at T$_{c,R}$. In contrast, the SAW induced dc signal peaks at T$_{c,m}$ 
(see figure \ref{magnetic_electric_A}).\\ 

This leads us to assume the second proposed mechanism: the traveling strain-wave modulates the local density, B$_v$, 
of vortices when they are pinned to the crystal lattice as shown for bulk samples with bulk ultrasound by Pankert et al.\cite{Pankert90}. 
This modulation of the local vortex density leads to the flow of local currents because of the Lorentz interaction. While the
origin of the induced dc-voltage is the ac-dc-conversion effect as described above, the origin of the ac-currents
differs and the induced dc-voltage drop is consequently related to the {\em magnetic} pinning transition
(the irreversibility line) and therefore leads to the observed 'background' dc voltage drop peaking at T$_{c,m}$
as determined by ac susceptibility experiments.\\
This signal does not depend on the sign of the applied magnetic field, since the time-averaged vortex motion vanishes.
Third, losses at the conversion of the rf-driving signal to SAW might slightly heat the sample
and lead to a temperature gradient in the superconducting film. The resulting Nernst signal,
however, is expected also at temperatures above T$_c$ and (at the small magnetic fields 
applied here) would be below the limits of the resolution of the experiment. Furthermore, the Nernst contribution
was ruled out by high-frequency amplitude-modulated experiments. The induced
dc-voltage was observed in lock-in experiments up to 100kHz, therefore a contribution of the Nernst effect,
the response time of which is limited to some 100ms by the thermal capacity of the sample, can be excluded.\\
\end{sloppypar}

\section{conclusions}
The influence of a surface acoustic wave on the motion of the magnetic vortex ensemble was determined by
analyzing the SAW-induced dc voltage perpendicular to the sound propagation direction. A pronounced
temperature dependent dc-voltage peak, which is antisymmetric in the applied magnetic field
strength, arises as consequence of SAW-induced directed vortex drag. This possibility of manipulating the
vortex motion and vortex density by a dynamic external parameter has far reaching technological
consequences. The introduction of standing SAWs and crossed SAW-fields is under investigation for
possible applications as flux-conveyors, flux-valves and switches.


\end{document}